\documentclass[pra,twocolumn]{revtex4-2}
  \usepackage{graphicx}
  \usepackage{amsmath}
  \usepackage{amssymb}
  \usepackage{makeidx}
  \usepackage{amsfonts}
  \usepackage{appendix}
  \usepackage{hyperref}
  \usepackage{mathrsfs}

  \hypersetup
  {
    colorlinks,%
    citecolor=black,%
    linkcolor=black,%
    urlcolor=black,%
  }

  \newcommand{\ket}[1]{\left| #1 \right\rangle}

\makeindex

\begin{document}

\title {Quantum teleportation of hybrid qubits and single-photon qubits using Gaussian resources}

\author{Soumyakanti Bose}
\email{soumyabose@snu.ac.kr}
\affiliation{Department of Physics $\&$ Astronomy, Seoul National University, 1 Gwanak-ro, Gwanak-gu, Seoul 08826, Korea}
\author{Hyunseok Jeong}
\email{h.jeong37@gmail.com}
\affiliation{Department of Physics $\&$ Astronomy, Seoul National University, 1 Gwanak-ro, Gwanak-gu, Seoul 08826, Korea}

\date{\today}

\begin{abstract}
In this paper, we compare single-photon qubits and hybrid qubits as information carriers through
quantum teleportation using a Gaussian continuous-variable channel. A hybrid qubit in our study
is in the form of entanglement between a coherent state and a single photon. We find that hybrid
qubits outperform photonic qubits when coherent amplitudes of the hybrid qubits are as low as $\alpha\lesssim 1$, while photonic qubits yield better results for larger amplitudes. We analyze effects of photon
losses, and observe that the overall character of teleportation for different qubits remains the same
although the teleportation fidelities are degraded by photon losses. Our work provides a comparative
look at practical quantum information processing with different types of qubits.
\end{abstract}
\maketitle

\section{Introduction} 

Quantum teleportation  \cite{qt_bennett, qt_braunstein} provides a platform to transmit unknown quantum information at the cost of shared entanglement. It plays a central role in various quantum information processing tasks such as broadband communication \cite{qcomm_loock}, quantum computing  \cite{qcomp_gottesman, qcomp_raussendorf, qcomp_knill}, and secret key distillation  \cite{qkd_braunstein}. 
Further developments provide feasibility of scalable quantum networks \cite{qnet_duan} leading to quantum internet \cite{qnet_kimble}. 
So far, quantum teleportation based on photonic systems has been experimentally demonstrated and extensively analysed for both discrete variable (DV) \cite{exp-tel-dv,dvtp_xia} and continuous variable (CV) \cite{exp-tel-cv,cvtp_braunstein} systems.  
While the teleportation of DV systems suffers from limited success probabilities \cite{dvtp_crit_lutkenhaus}, teleportation of CV states yields non-unit fidelities as it requires infinite squeezing that means an infinite amount of energy for perfect teleportation \cite{qt_braunstein}.

In order to circumvent the aforementioned difficulties, various attempts have been made using coherent-state qubits \cite{Enk01, Jeong01, JK02, JK02-com, Ralph03},
ancillary states \cite{Grice11,Ewert14}, squeezing operation \cite{Zaidi13,Kilmer19}, multiphoton qubits \cite{Lee15}, and hybrid qubits \cite{hbtp_appl_lee, hbtp_appl_kim}.
Another idea is to use Gaussian resources to teleport qubits \cite{hybrid-exp-2013, Men2014, hbtp_appl_hyung}. This approach has practical advantages that Gaussian quantum channels could readily be generated in a laboratory, and their characterizations are relatively straightforward \cite{gs_simon, gs_arvind}. 
On the other hand, it is more demanding to generate and analyze non-Gaussian channels \cite{ngs_gen_ralph, ngs_gen_shapiro, ngs_gen_shapiro2, ngs_gen_gea}. 
Despite earlier studies on teleporting various types of qubits, an analysis of comparative performance of these qubits via CV channels remains an open issue.


In this article, we provide a partial answer to this query. Here, we critically investigate the teleportation
of different types of qubits using an entangled Gaussian channel. 
To be specific, we focus on three different types of qubits: (a) the dual-rail single-photon qubit \cite{KLM}, (b) hybrid qubit of type A  that is entanglement between a single-photon state and a coherent state \cite{hyb-exp1} and (c) hybrid qubit of type B  that is entanglement between a single-photon state and superpositions of coherent sates~\cite{hyb-exp2}. 
As the entangled channel, we consider a pair of two-mode squeezed vacuum states (TMSVs), and analyze the performance of the CV-based teleportation  \cite{qt_braunstein}.

To that end, we first obtain analytic expressions for the teleportation fidelity and elaborate our observation through various plots. Our results indicate that in the
case of low amplitudes of the coherent states $\alpha \lesssim 1.0$ both types of the hybrid qubits, i.e. hybrid qubits (both types A and B) perform better than the single-photon qubit.
Hybrid qubits are known to be useful for fault-tolerant quantum computing with error correction  \cite{hbtp_appl_lee,omkar2020,omkar2021}, and the best suggested value of $\alpha$ for hybrid qubits (type A) is $\approx0.84$ \cite {omkar2021}. This suggests that  the teleportation scheme investigated in this paper may result in advantages in terms of fault-tolerant quantum computing.

On the other hand, when the amplitudes are $\alpha\approx 1.0$, all three states offer the same fidelities. 
Nonetheless, single-photon qubit outperforms  hybrid qubits of both types as the coherent amplitude increases ($\alpha \gtrsim 1.0$). 
Next, we consider a realistic case when the channel undergoes symmetric photon losses and suffer from decoherence. 
We observe that the photon loss degrades the performance for all three types of qubits so that quantum teleportation does not succeeds beyond a certain strength of loss. 
However, within the tolerable loss where quantum teleporation still prevails, the overall behaviour remains similar to the ideal cases, i.e., without loss.

The present article is organized as follows. In Sec.~\ref{sec_tf_pgs}, we generalize the Braunstein-Kimble (BK) protocol for continuous variable teleportation to the case of a four-mode resource and a two-mode input state. 
In Sec.~\ref{sec_qt_ideal}, we show the relative comparison between the single-photon qubit and the hybrid qubits under ideal conditions. 
Section~\ref{sec_qt_loss} involves similar comparisons under the effects of photon losses. 
In Sec.~\ref{sec_diss_conc}, we discuss various aspects of our results.

\section{Teleportation fidelity using pair-Gaussian channel}
\label{sec_tf_pgs}

In the original BK protocol for CV teleportation, which involves a bipartite resource and a single mode input, the fidelity F of teleportation could easily be evaluated in
terms of the characteristic functions as \cite{tfcf_chichov, tfcf_marian}
\begin{equation}
F=\int \frac{d^2 z}{\pi} \chi_{\rm{in}}(z) \chi_{\rm{res}}(z;z^*) \chi_{\rm{in}}(-z),
\end{equation}
where $\chi(z)=\rm{Tr}[\rho D(z)]$, $D(z)=\exp[za^\dagger - z^* a]$.
It is straightforward to extend this to a four-mode resource
and a two-mode input state as
\begin{align}
F&=\int \int \frac{d^2 z_1}{\pi} \frac{d^2 z_2}{\pi}~ \chi_{\rm{res}}(z_1,z_2;z_1^*,z_2^*) \times
\nonumber\\
&~~~~~~~~~~ \chi_{\rm{in}}(z_1,z_2) \times  \chi_{\rm{in}}(-z_1,-z_2).
\label{eq_fid_fm}
\end{align}
Here, we consider the four-mode Gaussian channel to be a pair of TMSV states described by the variance matrix $\Sigma=V\oplus V$, and $V$ stands for the variance matrix of the TMSV given by
\begin{equation}
V=\begin{pmatrix}
\eta &0 &c &0\\
0 &\eta &0 &-c\\
c &0 &\eta &0\\
0 &-c &0 &\eta
\end{pmatrix},
\label{eq_vm_tmsv}
\end{equation}
where $\eta=\cosh(2r)/2$ and $c=\sinh(2r)/2$.
As a consequence, the resource state characteristic function in Eq.~\eqref{eq_fid_fm} is
\begin{equation}
\chi_{\rm{res}}(z_1,z_2;z_1^*,z_2^*)= e^{-\vec{Z}^T \Sigma_c^{-1} \vec{Z}},
\label{eq_cf_pgs}
\end{equation}
where $\vec{Z}=(z_1,z_1^*,z_2,z_2^*)^T$ and $\Sigma_c=\Sigma_c^1 \oplus \Sigma_c^2$ s.t. $\Sigma_c^i=\begin{pmatrix}
0 & \eta-c\\
\eta-c & 0
\end{pmatrix}$ ($i=1,2$).
Replacing $\chi_{\rm{res}}(z_1,z_2;z_1^*,z_2^*)$ in \eqref{eq_fid_fm}, the fidelity for two-mode input with pair-Gaussian resource can be represented as
\begin{equation}
F=\int \frac{d\vec{Z}}{\pi^2} e^{-\vec{Z}^T \Sigma_c^{-1} \vec{Z}} \chi_{\rm{in}}(\vec{Z}) \chi_{\rm{in}}(-\vec{Z})
\label{eq_fid_pg}
\end{equation}
We then explore quantum teleportation with a pair-Gaussian channel for the aforementioned types of qubits.
It is known that in the case of qubits, quantum teleportation is successful when the fidelity is over the classical
limit, {\it i.e.}, $F>2/3$ \cite{qtnonlocal_horodecki, qtnonlocal_gisin}.

\section{Teleportation of different types of qubits with pair-Gaussian channel}
\label{sec_qt_ideal}
In this article, we consider three different types of qubits. They are (a) dual-rail single-photon qubit (spq), (b) hybrid-qubit type A (hqA) and (c) hybrid-qubit type
B (hqB) that can expressed as
\begin{eqnarray}
\ket{\psi_{\rm{spq}}} &=\sqrt{p}\ket{0,1} + \sqrt{1-p}~e^{i\phi}\ket{1,0}, \label{spq}\\
\ket{\psi_{\rm{hqA}}} &=\sqrt{p}\ket{0,\alpha} + \sqrt{1-p}~e^{i\phi}\ket{1,-\alpha}, \label{hbA}\\
\ket{\psi_{\rm{hqB}}} &=\sqrt{p}\ket{0,\alpha_{+}} + \sqrt{1-p}~e^{i\phi}\ket{1,\alpha_{-}},
\label{hbB}
\end{eqnarray}
where $\alpha_{\pm}=(\ket\alpha \pm \ket{-\alpha})/N_{\pm}$
are even ($\alpha_+$) and odd ($\alpha_-$) superpositions of coherent states with $N_{\pm}=2(1\pm e^{-2\alpha^2})$. 
It should be pointed out that hybrid qubits in the forms of Eq.~(\ref{hbA}) and Eq.~(\ref{hbB}) have been experimentally implemented in Ref.~\cite{hyb-exp1} and Ref.~\cite{hyb-exp2}, respectively.
These types of hybrid entanglement are useful for quantum computation \cite{hbtp_appl_lee,omkar2020,omkar2021} as well as quantum communication \cite{hbtp_appl_park, jbc16, hkim16, choi20}.
The input states in Eqs.~(\ref{spq}-\ref{hbB}) are parametrized by two parameters $p$ and $\phi$. 
We thus write $F$ in Eq.~(\ref{eq_fid_pg}) as $F(p,\phi)$ in the present context.
From an experimental point of view, it is natural to assume that parameters $p$ and $\phi$ would vary during the trials.
It is then imperative to look at the average behaviour of the teleportation process. 
This could be accomplished by considering the fidelity averaged over all possible input states, i.e.,
\begin{equation}
F^{\rm{av}}=\frac{1}{2\pi}\int dp\int d\phi~ F(p,\phi).
\label{eq_avfid}
\end{equation}
Henceforth, we shall refer to Eq.~\eqref{eq_avfid} only while talking
about fidelity, unless mentioned otherwise. Here we obtain an analytic expression for the teleportation fidelities for the input qubits as
\begin{widetext}
\begin{align}
F^{\rm{av}}_{\rm{spq}} &=4\frac{(1-2\beta)^2 + 4\beta^2}{(2+\Delta)^2},
\label{eq_avfid_eq}
\\
F^{\rm{av}}_{\rm{hqA}} &=\frac{4}{3(2+\Delta)^4} \left( (2+\Delta)^2 + (4+\Delta^2) + (2+\Delta)\left[ \Delta e^{-f(\alpha)} + 2e^{-4\alpha^2}e^{f(\alpha)} \right] \right),
\label{eq_avfid_hqA}
\\
F^{\rm{av}}_{\rm{hqB}} &=\frac{8}{3(2+\Delta)^2} \left( \left[ \frac{1}{N_+^2} + \frac{1}{N_-^2} \right]\left[ (1+e^{-4\alpha^2}) + (e^{-\frac{8\alpha^2}{2+\Delta}} + e^{-\frac{4\Delta \alpha^2}{2+\Delta}}) \right] - 4\left[ \frac{1}{N_-^2} - \frac{1}{N_+^2} \right]e^{-2\alpha^2} \right.
\nonumber
\\
&~~~~+~ \left. \frac{1}{N_+N_-}\left[ (1-e^{-4\alpha^2}) - \frac{2-\Delta}{2+\Delta}(e^{-\frac{8\alpha^2}{2+\Delta}} - e^{-\frac{4\Delta \alpha^2}{2+\Delta}}) \right] \right),
\label{eq_avfid_hqB}
\end{align}
\end{widetext}
where $\Delta=4(\eta -c)$, $\beta=1/(2+\Delta)$, $f(\alpha)=8\alpha^2/(2+\Delta)^4$, and
subscripts spq, hqA and hqB represent single-photon qubit, hybrid qubit of type A and  hybrid qubit of type B, respectively.
 
\begin{figure}[t]
\includegraphics[scale=0.7]{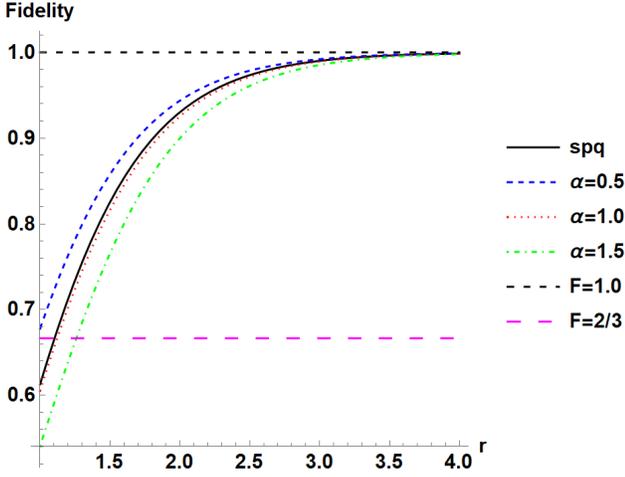}
\caption{Teleportation fidelities of single-photon qubit and
hybrid qubit of type A without photon loss. Here, `spq'  for the solid curve represents the single-photon qubit.
}
\label{fig_comp_eqhqI_ideal}
\end{figure}

In Figs.~\ref{fig_comp_eqhqI_ideal} and \ref{fig_comp_eqhqII_ideal}, we plot the comparison of quantum teleportation fidelities for the single-photon qubit
and the hybrid qubits with the pair-TMSV channel under the ideal condition without photon loss. As shown in the figures, hybrid qubits perform better as long as
the coherent amplitude is $\alpha < 1$.
When each type of qubit contains the average photon numbers of $\alpha \approx 1$,
there is little distinction between the different types of
qubits. With the increase in amplitude of the coherent
state ($\alpha \gtrsim 1$) the single-photon qubit appears to yield
better performance. It should also be noted that for all
three types of qubits, quantum teleportation is achieved
for the degree of squeezing is $r\gtrsim 1.0$.

\subsection{Comparison with the standard BK protocol using single two-mode Gaussian channel  for single-mode qubits}

\begin{figure}[t]
\includegraphics[scale=0.85]{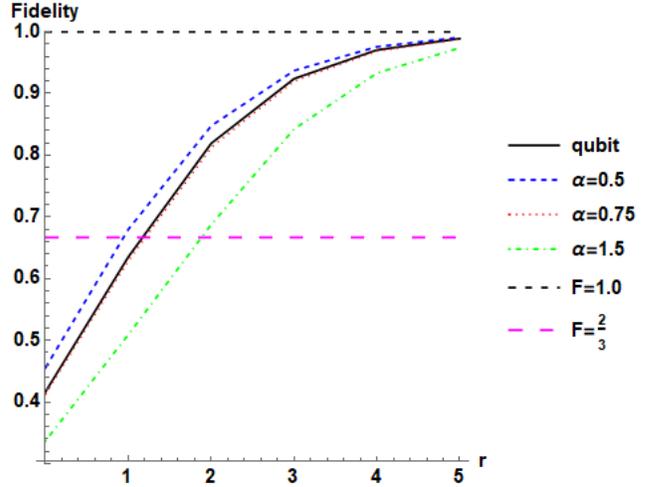}
\caption{Teleportation fidelities of single-mode single-photon qubit (a superposition of vacuum and single photon that is often called single-rail-logic qubit) and coherent-state qubits in the standard BK protocol. Here, `qubit' for the solid curve means the single-mode single-photon qubit $|\psi_0\rangle$.}
\label{fig_comp_pqcs_single}
\end{figure}

In order to apprehend the characteristics of the qubit teleportation using pair-Gaussian channel, we further consider the standard BK protocol for teleportation of single-mode qubits using a TMSV as the channel.
In Fig.~\ref{fig_comp_pqcs_single}, we plot the average fidelity for a single-mode single-photon qubit of $|\psi_0\rangle=\sqrt{p}\ket 0 + \sqrt{1-p}e^{i\phi}\ket 1$  \cite{sr1,sr2} and for coherent-state qubits of $|\psi_\alpha\rangle=(\sqrt{p}\ket\alpha + \sqrt{1-p}e^{i\phi}\ket{-\alpha})/N$ \cite{cat1,cat2,cat3,cat4} with normalization factor  $N=1+2\sqrt{p(1-p)} e^{-2\alpha^2} \cos\phi$. 
In a straightforward calculation using Eq.~\eqref{eq_avfid}, we find that the average fidelity for the single-mode single-photon qubit $|\psi_0\rangle$ is given by 
\begin{equation}
    F^{\rm{av}}_{0}=\frac{2}{3(2+\Delta)}\left( 2+\frac{4+\Delta^2}{(2+\Delta)^2} \right).
\end{equation}
Similarly, the average fidelity for the a coherent-state qubit ($|\psi_\alpha\rangle$) is obtained  as
\begin{eqnarray}
    F^{\rm{av}}_{\alpha}
    &=\frac{2}{\pi N^2 (2+\Delta)} \Big[ p^2 + (1-p)^2 + 4\sqrt{p(1-p)} e^{-\alpha^2}\cos\phi
    \nonumber\\
    &~+ 2p(1-p) \left( e^{-\frac{8\alpha^2}{2+\Delta}} + e^{-4\alpha^2}\cos(2\phi) +e^{-\frac{4\Delta\alpha^2}{2+\Delta}} \right) \Big].
    \nonumber\\
\end{eqnarray}

\begin{figure}[t]
\includegraphics[scale=0.7]{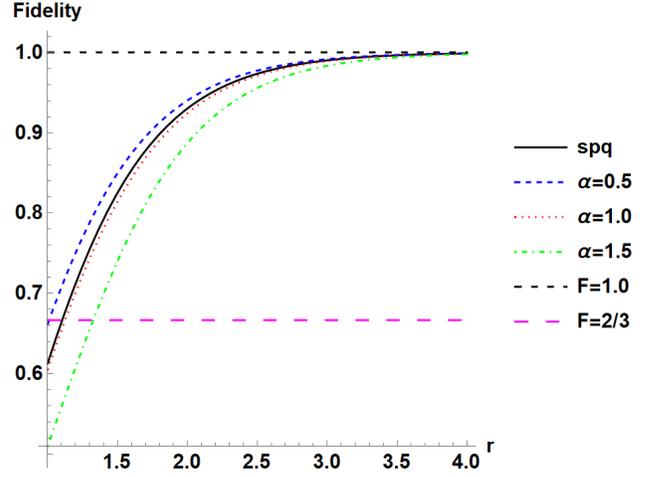}
\caption{Teleportation fidelities of single-photon qubit and
hybrid qubit of type B without photon loss.}
\label{fig_comp_eqhqII_ideal}
\end{figure}

\begin{figure}[t]
\includegraphics[scale=0.75]{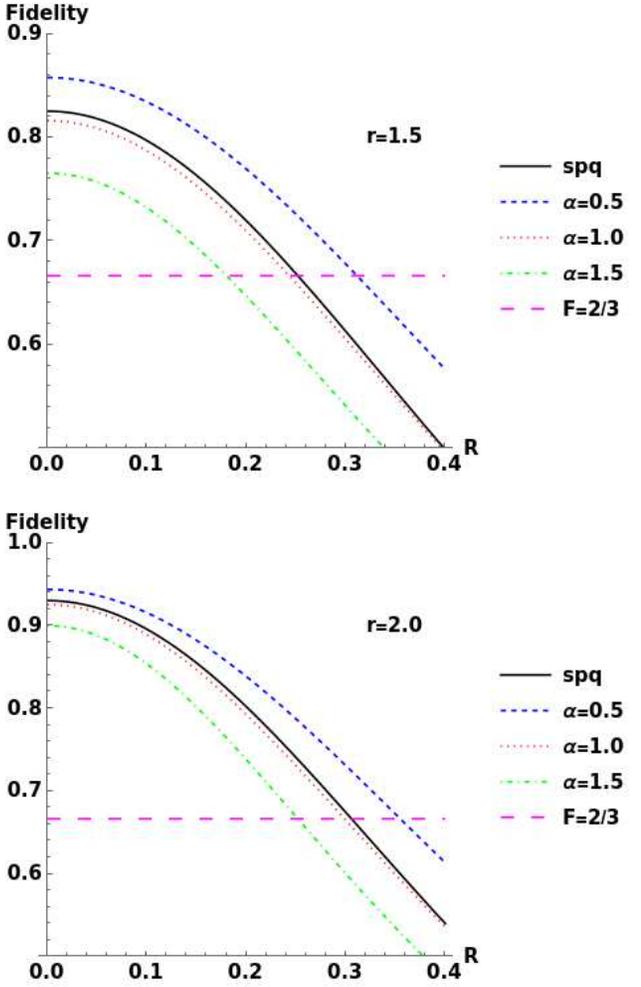}
\caption{Teleportation fidelities  for single-photon qubit and
hybrid qubit type A under photon loss. Here, $R$ is the loss parameter modeled by the refectivity of a beam splitter. 
}
\label{fig_comp_eqhqI_loss}
\end{figure}

\begin{figure}[t]
\includegraphics[scale=0.7]{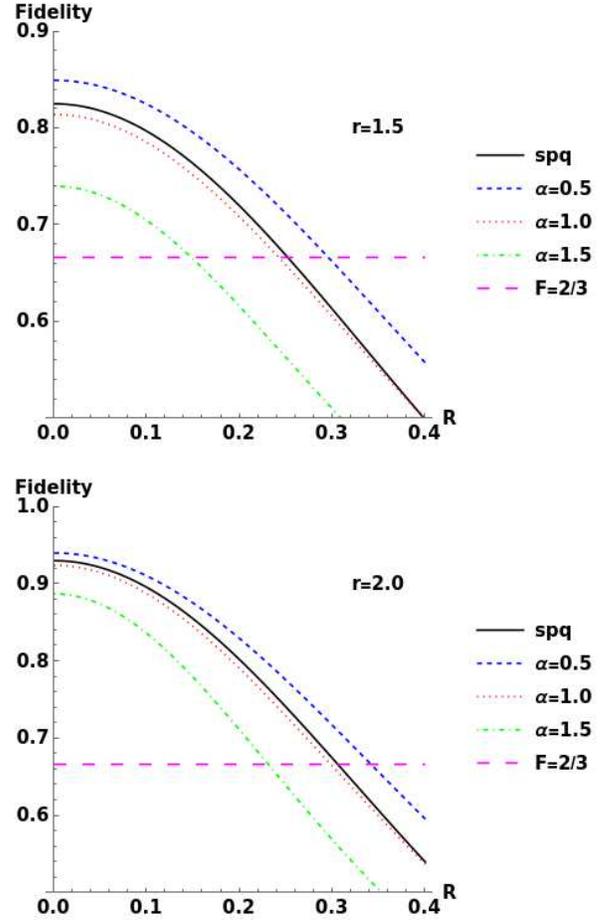}
\caption{Teleportation fidelities for single-photon qubit and
hybrid qubit type B under photon loss.}
\label{fig_comp_eqhqII_loss}
\end{figure}

As shown in Fig.~\ref{fig_comp_pqcs_single}, there are explicit distinctions between the parameter regions of $\alpha$ as observed in the case of coherent-state qubits. 
While $\alpha=0.5$  coherent-state qubits could be teleported more efficiently compared to the single-mode single-photon qubit. 
Similarly, for $\alpha \gtrsim 1$ the single-mode single-photon qubit outperforms the coherent-state qubits.
However, the original BK protocol and the extended BK protocol investigated in this paper yield somewhat different results when the single-mode single-photon qubit and the coherent-state qubit  show equivalent performances. I.e., as shown in Fig.~\ref{fig_comp_pqcs_single}, the two different types of qubits under the original BK protocol result in the equal fidelities for $\alpha\approx 0.75$ in contrast to the case with the extended BK protocol presented in Figs.~\ref{fig_comp_eqhqI_ideal} and \ref{fig_comp_eqhqII_ideal} where $\alpha\approx1$ is such a boundary value.

\section{Teleportation of qubits using pair-Gaussian channel under photon loss}
\label{sec_qt_loss}

In this section we consider teleportation of the qubits under decoherence. 
For the sake of simplicity we consider that the resource, i.e., the pair-Gaussian channel undergoes symmetric photon loss. 
This could be modeled as followed. First each the modes of the Gaussian state passes through one of the input modes of two passive beam splitters (BS) of refectivity $R$ while other inputs are kept at vacuum. 
Subsequently discarding the output ancilla modes (taking trace) leads to the effective channel under photon loss. 
Here, $R = 1$ stands for complete photon loss. 
The input-output mode relation for this BS could be written as
\begin{equation}
\begin{pmatrix}
a_{\rm{out}} \\
b_{\rm{out}}
\end{pmatrix}
=
\begin{pmatrix}
\sqrt{1-R} & \sqrt{R}\\
-\sqrt{R} & \sqrt{1-R}
\end{pmatrix}
\begin{pmatrix}
a_{\rm{in}} \\
b_{\rm{in}}
\end{pmatrix}.
\end{equation}
In a straightforward calculation the effect of this symmetric photon loss could be represented as the variance matrix of a TMSV in Eq.~\eqref{eq_vm_tmsv} changing to
\begin{equation}
V_{\rm{loss}}=\begin{pmatrix}
\eta^{'} &0 &c^{'} &0\\
0 &\eta^{'} &0 &-c^{'}\\
c^{'} &0 &\eta^{'} &0\\
0 &-c^{'} &0 &\eta^{'}
\end{pmatrix},
\label{eq_vm_tmsv_loss}
\end{equation}
where $\eta^{'}=\frac{1+2(1-R)\sinh^2 r}{2}$ and $c^{'}=(1-R)\cosh r \sinh r$.
Corresponding fidelities in Eqs.~\eqref{eq_avfid_eq}, \eqref{eq_avfid_hqA} and \eqref{eq_avfid_hqB} could be accordingly obtained by replacing $\eta$ and $c$ by the respective primed quantities \eqref{eq_vm_tmsv_loss}.

In Figs.~\ref{fig_comp_eqhqI_loss} and \ref{fig_comp_eqhqII_loss}, we plot teleportation fidelities for the single-photon qubit and hybrid qubits under symmetric photon losses. For comparison, we consider two different
squeezing values of $r=1.5$ and $2.0$.
We observe that the overall performances of the qubits remain the same except for the fact that the loss parameter ($R$) now restricts the experimentally available region over which teleportation is performed over the classical limit. As evident from the figures, quantum teleportation does not work beyond $R \sim 0.35$.

\section{Conclusion}
\label{sec_diss_conc}

We have analysed quantum teleportation of single-photon qubits and hybrid qubits using a pair of Gaussian channels. 
We have considered two slightly different types of hybrid qubits. 
One is entanglement between a single photon and a coherent state, and the other is entanglement between a single photon and a coherent-state
superposition. 
We note that hybrid qubit of both types considered in this paper have been experimentally implemented \cite{hyb-exp1,hyb-exp2} although their fidelities are yet limited.
Our analysis is mostly centred around the query of which type of qubit is better as a information carrier when a Gaussian channel is used for quantum teleporation.

To that end, we have shown that while the coherent amplitude is as low as $\alpha\lesssim 1$, both types of hybrid qubits yield better results than the single-photon qubit. 
On the other hand, as $\alpha$ increases, single-photon qubits lead to better performance. 
This difference could be explained in terms of non-Gaussian characters of the states. 
Since we have considered teleportation of non-Gaussian states using a Gaussian channel, it is quite natural that an increase in the non-Gaussianity of the input states will mar the performance of the teleportation. 
With increase in $\alpha$, non-Gaussianity of the hybrid states increases which, in turn, brings forth the drop in the corresponding performance.

We have further analyzed the performance of qubit teleportation under symmetric photon loss effects. Our analytic results indicate that under decoherence, the overall characteristics remain the same; hybrid qubits work better for relatively small amplitudes. 
Of course, the loss restricts the parameter region over which quantum teleportation is performed over the classical limit. 
In other words, higher loss requires higher squeezing for valid quantum teleportation.


Our work provides a comparative look at practical quantum information processing with different types of qubits.
Hybrid qubits are useful for various applications including quantum computation \cite{hbtp_appl_lee,omkar2020,omkar2021} and quantum communication \cite{sheng13, lim16, parker17, parker20, wen-arx}.
It is particularly advantageous for fault-tolerant quantum computing with error correction  \cite{hbtp_appl_lee,omkar2020,omkar2021}. 
The best suggested value of $\alpha$ for hybrid qubits (type A) to perform fault-tolerant quantum computing was found to be $\approx0.84$ \cite {omkar2021}. 
This renders immediate importance and relevance to our work in the context of practical quantum information processing with hybrid systems.

\end{document}